\documentclass[twocolumn,epjc3]{svjour3}  
\smartqed  
\RequirePackage{graphicx}
\usepackage[utf8x]{inputenc}
\usepackage{amsmath}
\usepackage{amssymb}
\usepackage[section]{placeins}
\usepackage{units}
\usepackage{rotating}
\usepackage{cite}

\newcommand{\kt}{k_\perp}
\newcommand{\qt}{q_\perp}
\newcommand{\pt}{p_\perp}
\newcommand{\qhat}{\hat q}

\newcommand{\der}[2][]{\frac{\d#1}{\d#2}}

\renewcommand{\d}{\text{d}}

\journalname{Eur. Phys. J. C}

\begin{document}


\title{Coherent Radiative Parton Energy Loss beyond the BDMPS-Z Limit}

\author{Korinna Christine Zapp\thanksref{email1,addr1} \and Urs Achim Wiedemann\thanksref{addr2}}

\thankstext{email1}{e-mail: k.c.zapp@durham.ac.uk}

\institute{Institute for Particle Physics Phenomenology, Durham University, Durham\ DH1\,3LE, UK \label{addr1}
  \and Department of Physics, CERN, Theory Unit, CH-1211 Geneva 23\label{addr2}}

\date{Received: date / Accepted: date / }

\maketitle

\begin{abstract}

It is widely accepted that a phenomenologically viable theory of jet quenching for heavy ion
collisions requires the understanding of medium-induced parton energy loss beyond the limit
of eikonal kinematics formulated by Baier-Dokshitzer-Mueller-Peign\'e-Schiff and Zakharov (BDMPS-Z). 
Here, we supplement a recently developed exact Monte Carlo implementation of the BDMPS-Z
formalism with elementary physical requirements including exact energy-momentum conservation,
a refined formulation of jet-medium interactions and a treatment of all parton branchings on the same
footing. We document the changes induced by these physical requirements and we describe their kinematic origin.
\end{abstract}

\section{Introduction}
Over the last decade, experiments at RHIC~\cite{Adcox:2004mh,Back:2004je,Arsene:2004fa,Adams:2005dq} 
and at the LHC (see preliminary data in ~\cite{QM11})
have demonstrated that the fragmentation
pattern of highly energetic partons ($E_\perp \gtrsim \unit[10]{GeV}$) is altered strongly when embedded in the
dense QCD matter produced in ultra-relativistic nucleus-nucleus collisions. This is seen
in particular in the inclusive high-$\pt$ hadronic  spectra that are strongly suppressed 
by a factor up to 5 (7) at RHIC (LHC) compared to expectations from proton-proton
spectra~\cite{Adams:2005ph,Adare:2008qa,Aamodt:2010jd,cmsraa}, and that stay suppressed over the entire  
high-$\pt$ range studied experimentally so far 
(up to $\pt \sim \unit[100]{GeV}$ at the LHC). Additional information about this jet quenching phenomenon
comes from a broad range of jet-like particle correlation measurements~\cite{QM11}, and most recently from the
observation of  strong modifications on the level of reconstructed jets in nucleus-nucleus
collisions~\cite{Aad:2010bu,Chatrchyan:2011sx}, where first steps towards a characterization
 of the entire medium-modified jet fragmentation pattern have been taken. 

In general, both inelastic (giving rise to radiative energy loss) and elastic processes (giving rise to collisional
energy loss) are expected to contribute to the strong modification of the internal structure
of the parton shower, leading in particular to an energy degradation of the most 
energetic partons as well as effects on transverse momentum broadening and intra-jet multiplicity. 
However, the QCD-based analytical analysis of jet quenching remains restricted so far to 
kinematical limiting cases. In particular, studies of the dominant inelastic process of 
medium-induced gluon radiation have focussed so far on an eikonal high-energy 
approximation~\cite{Baier:1996sk,Zakharov:1997uu,Wiedemann:2000za,Gyulassy:2000er,Wang:2001ifa,Arnold:2002ja}, 
according to which
the energy of the projectile parton $E$ is taken to be much larger than the energy of the radiated
gluon $\omega$, which in turn is treated in the collinear approximation with transverse  momenta carried
by the gluon ($\kt$) and the recoiling scattering center ($\qt$) much smaller than the gluon energy, 
\begin{equation}
 E \gg \omega \gg \kt,\, \qt \gg \Lambda_\text{QCD} \,.
\label{eq::bdmps}
\end{equation}
The seminal analyses of radiative parton energy loss by Baier, Dokshitzer, Mueller, Peign\'e, Schiff 
~\cite{Baier:1996sk} and by Zakharov~\cite{Zakharov:1997uu} (BDMPS-Z) are based on this 
approximation. This also applies to more recent analytical formulations (for an overview,
see Ref.~\cite{Armesto:2011ht}) and to those Monte Carlo 
models~\cite{Deng:2010mv,Armesto:2009fj,Schenke:2009gb,Lokhtin:2008xi,Zapp:2008gi} that aim at implementing 
the analytically known QCD-based calculations.

However, jet quenching phenomenology requires 
an understanding of the interaction between partonic projectile and QCD medium beyond the kinematic
range (\ref{eq::bdmps}). It is generally agreed that extrapolating calculations of parton energy loss from 
  (\ref{eq::bdmps}) to the full phenomenologically relevant kinematical range induces
  uncertainties that are much larger than other known model-dependent differences~\cite{Armesto:2011ht}. 
 In addition, the analytical calculations of parton energy loss based on  (\ref{eq::bdmps}) show qualitative
 features that are physically sensible only within this reduced kinematical range of validity. For instance,
 they conserve energy and momentum only up to corrections of order $O(\omega/E)$ and $O(\kt/\omega)$,
 an approximation that is justified for the range (\ref{eq::bdmps})  but that can introduce large uncertainties in 
 phenomenologically relevant kinematic regimes. An analogous conclusion applies to the treatment of recoil
 effects; collisional energy loss is negligible in the kinematics of (\ref{eq::bdmps}) but may be
 sizeable in phenomenologically relevant kinematic regimes. Furthermore, 
 the strong ordering of the energy fractions $E-\omega \gg \omega$ is an obstacle for treating all
 daughter partons subject to the same medium-modified dynamics. Also, within existing analytical treatments,
 the possibilities of studying the dependence of parton energy loss on properties of the medium is limited.
 For instance, it is difficult to vary the composition of the medium in terms of elastic and inelastic scattering
 centers, their hardness and their energy dependence.
 
Many of the above-mentioned limitations arise from the use of analytical techniques and 
 could be avoided if one formulated the dynamics of jet quenching in terms of a Monte Carlo algorithm. 
 For instance, exact energy-momentum conservation or the democratic treatment of all partonic splittings
  are implemented easily in Monte Carlo approaches while their inclusion in analytical formulations is
 complicated. Also, Monte Carlo techniques enhance naturally the versatility in testing different scattering 
 properties of the medium or in interfacing with different hadronisation models. In practice, 
 if one wants to improve the applicability of a calculational framework with Monte Carlo techniques, it is an 
 obvious prerequisite to establish with which accuracy and in which kinematic range a Monte Carlo
 algorithm reproduces by construction a defined calculational framework. With this motivation, we have
 established recently a MC algorithm that faithfully accounts for the dominant medium-induced quantum
 interference effects present in the BDMPS-Z formalism~\cite{Zapp:2008af,Zapp:2011ya}. In the present 
 work, we document how results of this MC version of the BDMPS-Z formalism are altered, when the MC simulation outside
the kinematic range (\ref{eq::bdmps}) is supplemented by various physical requirements.
  
\section{Extending the MC algorithm beyond the BDMPS-Z limit}
\label{sec::steps}

In the following, the setting in which the MC algorithm reproduces quantitatively all features of the BDMPS-Z 
formalism will be referred to as BDMPS-Z limit. This algorithm is described in detail in ~\cite{Zapp:2011ya}.
In the BDMPS-Z limit, the medium is simulated as
a source of elastic and inelastic scattering centres, characterized by  the mean free paths $\lambda_\text{inel}$ 
and $\lambda_\text{elas}$ and the corresponding differential cross sections
\begin{equation}
 \der[\sigma_\text{elas}]{\mathbf{q}_\perp} \propto C_\text{R} \frac{1}{(\mathbf{q}_\perp^2 + \mu^2)^2}
\theta(2\mu - |\mathbf{q}_\perp|)\, ,
\label{elscatt}
\end{equation}
and
\begin{equation}
 \frac{\d \sigma_\text{inel}}{\d \omega\, \d \mathbf{q}_\perp\, \d \mathbf{k}_\perp} \propto C_\text{R}
\der[\sigma_\text{elas}]{\mathbf{q}_\perp} \frac{1}{\omega} \delta(\mathbf{k}_\perp-\mathbf{q}_\perp) \,. 
\end{equation}
With this choice, the BDMPS-Z transport coefficient that is typically used to characterize the medium reads
$\hat q_\text{MC} \equiv \mu^2/\lambda_\text{elas}$. Here, $\lambda_\text{elas} \equiv 1/n \sigma_\text{elas}$
and $n$ denotes the density of scattering centers. In the following, we study extensions of the BDMPS-Z 
baseline algorithm that account for the following physical requirements:
\begin{enumerate}
    \item {\it energy-momentum conservation}\\
		In the BDMPS-Z  limit (\ref{eq::bdmps}), the projectile parton energy remains
		unchanged by gluon emission. Also, the transverse momentum accumulated by the gluon is independent of
		the gluon energy. Here, we extend this MC algorithm to include exact local energy and 
		momentum conservation of all partonic interactions and splittings. 
    \item {\it refined formulation of jet-medium interactions}\\
        In the BDMPS-Z limit, transverse momentum transfers are bounded by $\qt \le 2\mu$ to ensure a 
        multiple soft scattering scenario. Here, we consider extensions that vary the scale and energy
        dependence of elastic and inelastic interactions with the medium. In particular
        \begin{enumerate}
        	\item {\it beyond the soft multiple scattering approximation}\\
	     The extended MC algorithm will allow for momentum transfers in the extended range $\qt \leq \omega$.
	     \item {\it energy-dependent cross section}\\
	      In general, the scattering cross sections can acquire an energy dependence, e.g. due to 
	      phase space restrictions. We consider a scenario with energy dependent mean free paths
		\begin{equation}
		 \lambda_\text{elas}(\omega) \propto \frac{\omega^2 + \mu^2}{\omega^2} 
		 \end{equation}
		and
		\begin{equation}
 		\lambda_\text{elas}(E_\text{proj}) \propto
		\frac{\lambda_\text{elas}(E_\text{proj})}{\ln\left( E_\text{proj}/\omega_\text{min} \right)}\, .
		\end{equation}
		As explained in Ref.~\cite{Zapp:2011ya}, the cut-off dependence of the MC algorithm on the 
		kinematic range $\left[ \omega_{\rm min}, E_{\rm proj}\right]$ of the gluon energy $\omega$
		does not add to the uncertainties of the BDMPS-Z limit, since it can be absorbed
		in a logarithmic dependence of the inelastic mean free path.
        \end{enumerate}
        
 \item {\it no overlap of formation times}\\
 The somewhat surprising finding of~\cite{Zapp:2011ya} was that in order to reproduce the BDMPS-Z limit 
 exactly, one has to allow for overlapping gluon formation times. Firstly, this complicates numerical implementations
 considerably. Secondly, it still remains to be shown whether the effect of overlapping formation times persists
 in more complete calculations of multiple medium-induced gluon radiation (for a first step towards such calculations, 
 see Refs.~\cite{MehtarTani:2010ma,CasalderreySolana:2011rz,MehtarTani:2011gf}).
 It is therefore interesting to quantify the difference between an exact implementation of 
 the BDMPS-Z results, and a simplified 'no overlap' implementation in which only one gluon can be formed at a time.
 
 \item {\it democratic treatment of all partons}\footnote{We restrict the study of this democratic treatment
  of all parton branchings to simulations in which exact energy-momentum
  conservation is implemented. The (partonic) multiplicity and the total energy of the ensemble would grow 
  unphysically large in a scenario in which energy-momentum conservation is treated only 
  approximately according to (\ref{eq::bdmps}) while all partons are subject  to the same dynamics
  of medium-induced rescattering and splitting.}\\
  In the BDMPS-Z limit, one neglects elastic interactions of partons
  with energy $O(E)$ (since elastic momentum transfers can be neglected in the kinematical range
  (\ref{eq::bdmps})), and one neglects inelastic branchings of radiated gluons (since the calculation
  focusses on the medium modifications of the most energetic parton)~\cite{Zapp:2011ya} . 
  Here, we include both. It is assumed that the new mean free paths can be approximated by adjusting the colour factors of the BDMPS-Z ones,
\begin{equation}
 \begin{array}{ll}
  \lambda_\text{elas}^\text{g} = \lambda_\text{elas}\, ,  & \qquad  \lambda_\text{elas}^\text{q} =
\frac{C_\text{A}}{C_\text{F}}\lambda_\text{elas}\, , \\
  \lambda_\text{inel}^\text{g} = \left( \frac{C_\text{F}}{C_\text{A}} \right)^2 \lambda_\text{inel}\, ,  & 
 \qquad \lambda_\text{inel}^\text{q} = \lambda_\text{inel} \, .
 \end{array}
\end{equation}
Scenarios without democratic treatment will be referred to as {\it standard} in the following. 
Subjecting all partons to the same medium-dependent dynamics may be regarded  
as a first step towards describing jet-like observables.
 It obviously influences the inclusive gluon spectrum and angular distribution, but it also provides
 additional information for a study of the suppression pattern of single-inclusive hadron spectra. 
 In particular, instead of calculating the energy loss of the "projectile parton" as in formalisms based on the 
 BDMPS-Z limit, one can now characterise  the energy fraction carried by the most energetic 
 partonic fragment in the ensemble, irrespective of whether this fragment is the projectile parton
 or some of the other (mostly gluonic) fragments. 
\end{enumerate}        
In the following, we discuss the impact of these model improvements on different observables.

\begin{figure}
   \centering \begin{turn}{-90} \includegraphics[width=.333\textwidth]{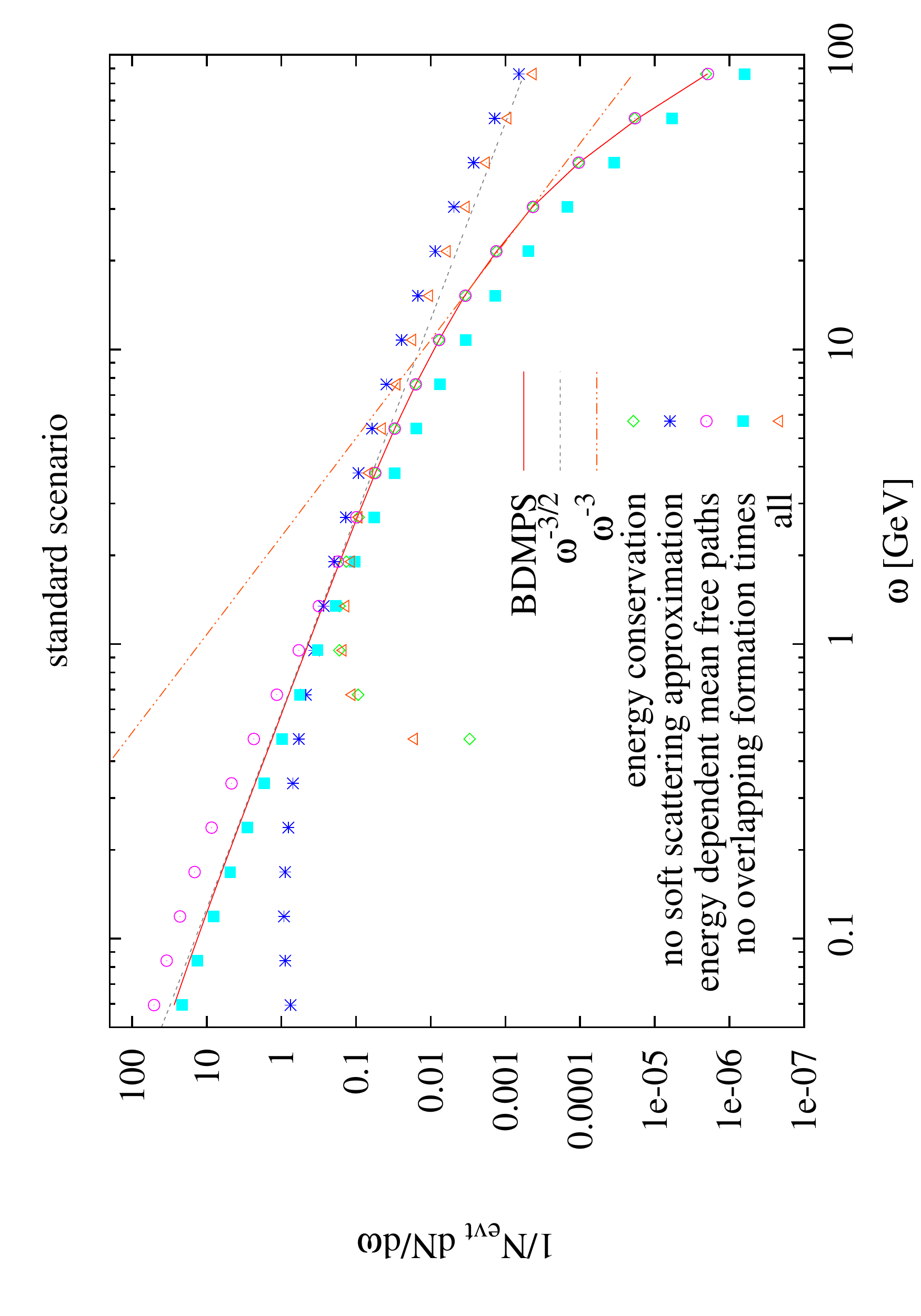} \end{turn}
   \centering \begin{turn}{-90} \includegraphics[width=.333\textwidth]{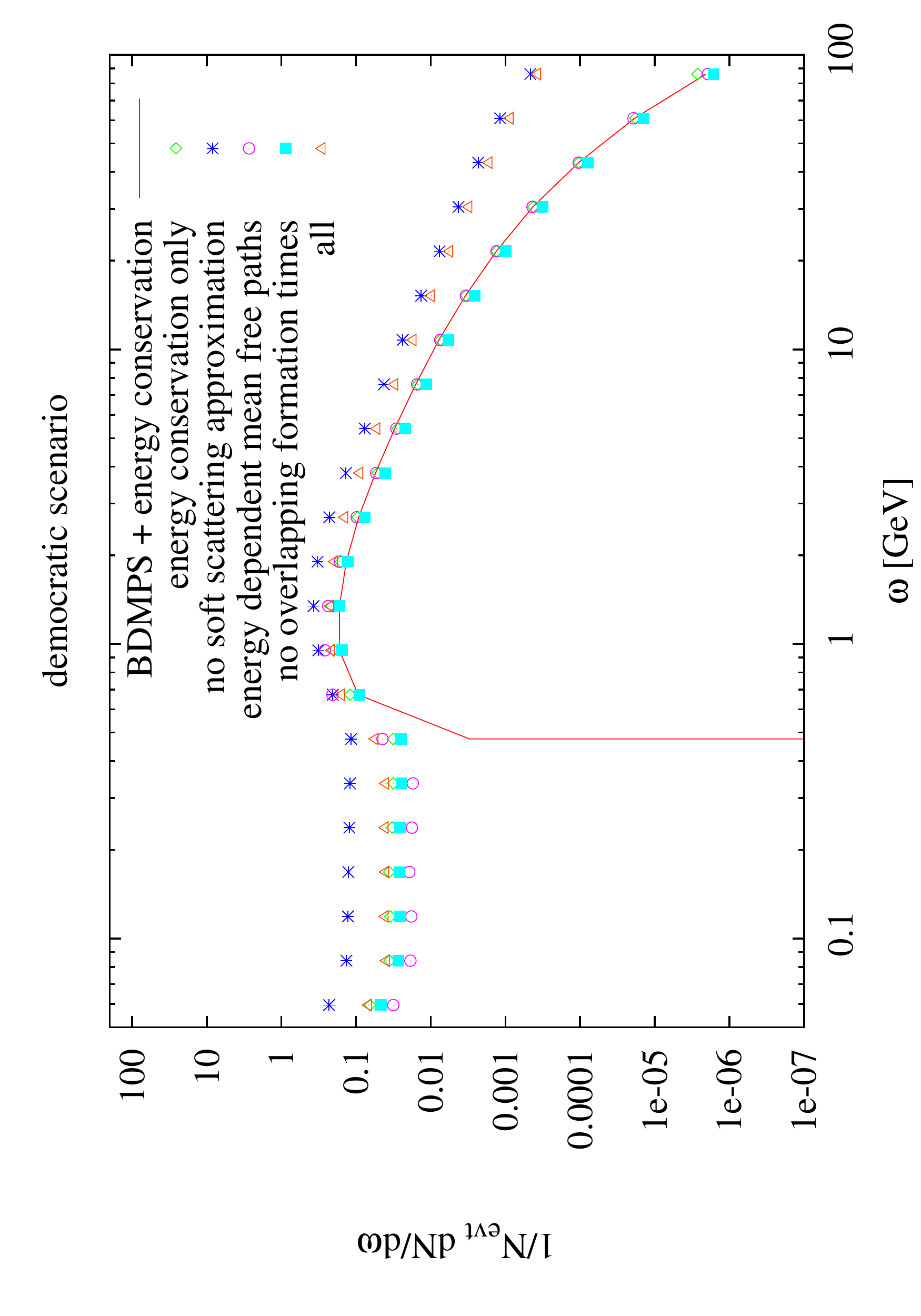} \end{turn}
   \caption{MC results for the gluon energy spectrum, shown are the BDMPS-Z scenario together with the modifications
explained in section~\ref{sec::steps} (parameters: $L=\unit[2.5]{fm}$, $E_\text{proj}=\unit[100]{GeV}$,
$\lambda_\text{elas}=\unit[0.1]{fm}$, $\lambda_\text{inel}=\unit[0.1]{fm}$, $\mu=\unit[0.7]{GeV}$,
$\omega_\text{min}=\unit[50]{MeV}$).}
   \label{fig::omega}
\end{figure}

\section{Gluon spectrum}
\label{sec3}

In the BDMPS-Z calculation the gluon spectrum has the characteristic dependence
\begin{equation}
 \der[I]{\omega} \propto \left\{ \begin{array}{ll}
                                  \omega^{-3/2} & \quad \text{for} \quad \omega \ll \omega_\text{c} \\
                                  \omega^{-3} & \quad \text{for} \quad \omega \gg \omega_\text{c}
                                 \end{array} \right.
 \label{eq::bdmpsspec} \,,
\end{equation}
where the characteristic gluon energy $\omega_\text{c} = \qhat L^2/2$ is the energy for which
the entire medium acts coherently. In Fig.~\ref{fig::omega}, we compare this analytical dependence
with results for the BDMPS-Z Monte Carlo baseline and for the extensions
discussed in the previous section. For the MC results in the BDMPS-Z limit, the spectrum shows the 
expected behaviour (\ref{eq::bdmpsspec}), albeit with a steeper fall-off towards the kinematic limit.

In general, enforcing energy and momentum conservation has a negligible effect on the energetic part of the gluon
distribution, since the radiation of several very energetic gluons in the same event is highly unlikely. The
soft part of the spectrum, on the other hand, is cut off completely due to momentum conservation. We understand this 
by observing that  the requirement $\kt \le \omega$ leads to a minimal formation time
\begin{equation}
 \tau = 3 \frac{2 \omega}{\kt^2} \ge 3 \frac{2}{\omega} = \tau_0 \, .
\label{eq::time}
\end{equation}
Since the formation time has to be smaller than the medium length $L$, gluons with energy smaller than 
$\omega_0 = 6/L$ cannot be radiated\footnote{The factor 3 in Eq.~\ref{eq::time} arises from the observation
that in QCD-based calculations of destructive interference terms, the phase has to become larger than $\sim 3$ in
order for the gluon to decohere from the projectile~\cite{Zapp:2011ya}.}.


If one allows radiated gluons to branch again ('democratic scenario'), the large-$\omega$ tail of the spectrum
remains unchanged since energetic gluons have long formation times, which inhibits further emissions. 
At very low energies,  however, the part of the spectrum that was inaccessible in the standard
scenario gets populated by gluons that moved into that region by radiating away parts of their energy,
see Fig.~\ref{fig::omega}.


Compared to the extensions of the MC algorithm discussed so far, lifting the soft scattering approximation leads to 
a quantitatively much more important increase in the number of energetic gluons. This is so, since 
a gluon can decohere much faster than in the soft scattering case, when a rather substantial 
transverse momentum transfer can be given in a single scattering.  In the standard scenario, the spectrum 
at low gluon energy is still suppressed, since the condition $\qt \le \omega$ is a more stringent bound than 
$\qt \le 2 \mu$, which is used in the BDMPS-Z simulation.
This leads to the suppression of low-energy gluons in the standard scenario. The shape of the gluon spectrum changes to
an 
approximate $\omega^{-3/2}$ dependence at large energies and is nearly flat for small energies.
In the democratic scenario the suppression of soft gluons is overcome by the enhanced radiation 
leading to an increase in the entire $\omega$-range.


Rescaling the mean free paths to take into account their energy dependence leads to a slight increase in radiation
around energies below about \unit[1]{GeV} (Fig.~\ref{fig::omega}), while requiring that only one gluon can be 
formed at a time leads to the expected suppression of gluon radiation over the
entire energy range. The size of the effect 
depends on the relation between the elastic and inelastic mean free paths and is only mild in this example.


In summary, the most dramatic changes in the gluon spectrum come from imposing energy and momentum conservation and from relaxing the soft scattering approximation. We note in this context that there 
is no a priori knowledge as to whether jet-medium interactions are dominated entirely by soft
momentum exchanges with $q_\perp < 2\mu$, or whether the $\propto 1/q_\perp^4$-tail of hard perturbative
processes makes a numerically significant contribution. Both cases 
may be regarded as corresponding to different, a priori conceivable properties of the dense QCD matter 
produced in heavy ion collisions. The strong sensitivity of the medium-induced gluon 
radiation pattern to such differences in the nature of medium-induced momentum transfers may be 
regarded as a phenomenologically wanted feature, as it points to the possibility of differentiating between 
different dynamical mechanisms of jet-medium interactions.

\section{Radiated energy}
\label{sec4}

\begin{figure}
  \centering
   \centering \begin{turn}{-90} \includegraphics[width=.333\textwidth]{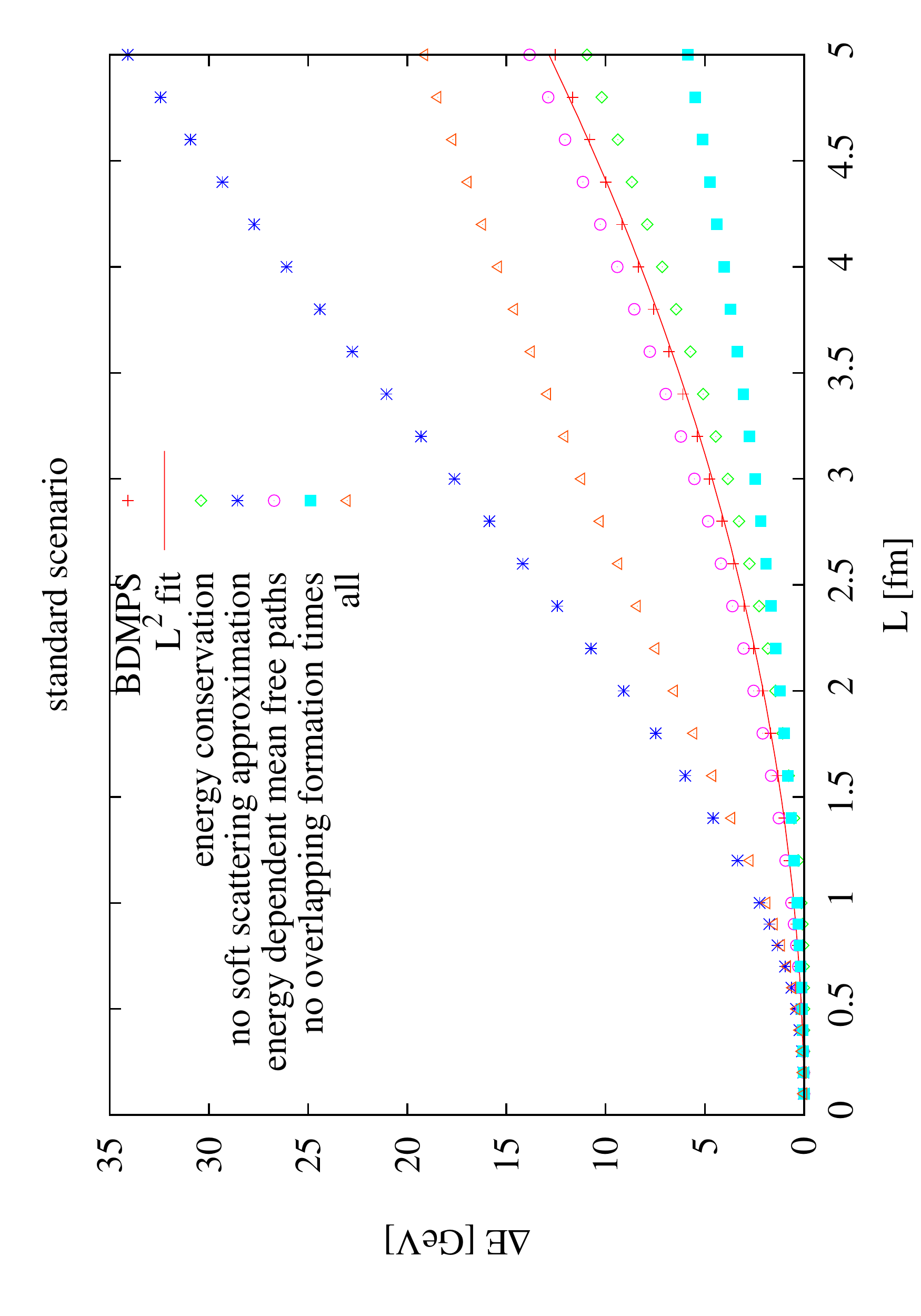} \end{turn}
   \centering \begin{turn}{-90} \includegraphics[width=.333\textwidth]{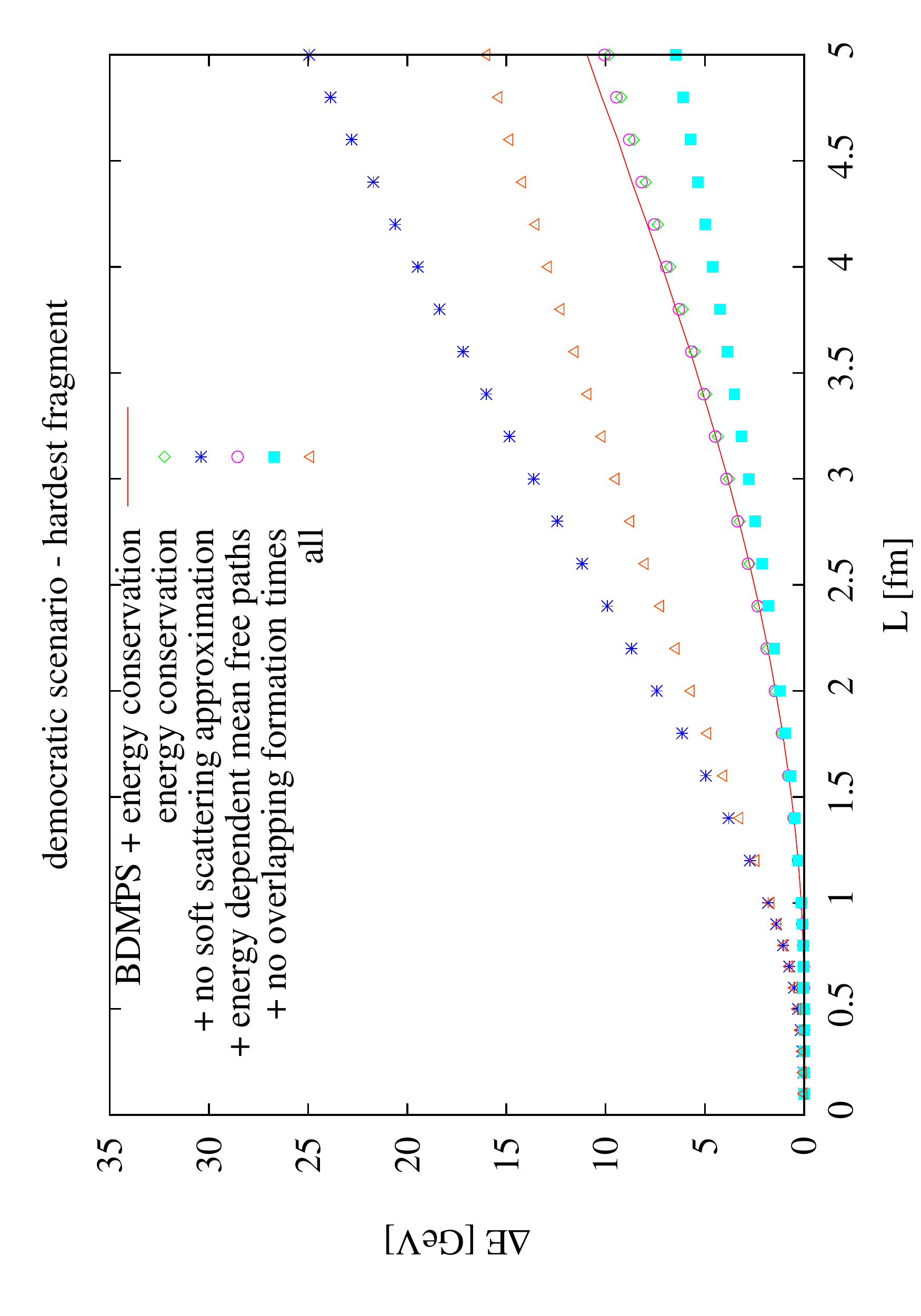} \end{turn}
   \caption{MC results for the radiative energy loss in the standard scenario (top), where the energy loss is defined
as the energy radiated away by the projectile quark, and democratic scenario (bottom), where the energy loss is defined
as the difference between the energy of the most energetic fragment and the incoming quark energy. Note that in the
lower panel the 'BDMPS + energy conservation' calculation has the standard definition of the energy loss (parameters as
in Fig.~\ref{fig::omega}).}
   \label{fig::deltae}
\end{figure}

In the BDMPS-Z limit, the radiative energy loss $\Delta E$ is uniquely defined as the energy radiated off the projectile
parton and is given by the first moment of the gluon spectrum. It has a quadratic path length dependence characteristic
of the
LPM-effect. As shown in Fig.~\ref{fig::deltae} this behaviour is reproduced to high accuracy by the MC algorithm.
In the standard scenario and for sufficiently small medium path length $L$, conservation of energy and momentum 
has only a mild effect on the energy loss, as mainly low energy emissions are affected. 
But for larger in-medium path length, deviations from the BDMPS-Z formalism due to four-momentum conservation
are more pronounced. In particular, for medium path lengths $L$ that exceed the typical 
coherence length $L_c$ of the most energetic gluons $\omega \sim E_\text{proj}$, the average parton 
energy loss is known to grow linearly in the BDMPS-Z formalism~\cite{Zapp:2008af,Zapp:2011ya},
while it is bound to $\Delta E < E_\text{proj}$ by exact energy momentum conservation. Here, we did 
not explore further this trivial and known effect that sets in only at very large in medium path lengths where 
it plays an important if not dominant role~\cite{Zapp:2008af}.

Relaxing the soft
scattering assumption, on the other hand, not only significantly increases the energy loss by enhancing energetic radiation, but also changes the path length dependence from quadratic to linear (except for very small $L$). 
The reason is that harder momentum transfers can destroy coherence leading to quasi-incoherent emission of energetic
gluons.

Energy-dependent mean free paths and no overlapping formation 
times lead to a slight enhancement and a somewhat bigger reduction of the energy loss, respectively, without 
changing the $L$-dependence much. When all these extensions of the MC algorithm are included, one finds
that the removal of the soft scattering approximation induces the most significant changes both in the 
$L$-dependence and the absolute magnitude of $\Delta E$. 

If one allows radiated partons to split again (democratic treatment), one can define the radiative energy loss 
alternatively as the difference between the energy
of the incident projectile parton and the energy of the most energetic parton at the end of the cascade. This
definition is arguably more physical, as a 
unique correspondence between the initial projectile parton and the most energetic final parton
exists only in the BDMPS-Z limit. (In general, the most energetic parton may very well be one of the gluons radiated off
the projectile quark.)
Obviously, the so defined energy loss cannot be larger than the energy radiated away by the projectile quark. 
This is clearly seen on the lower panel of Fig.~\ref{fig::deltae}. The energy loss is smaller, but the
qualitative picture of how extensions of the MC algorithm affect the result is  very similar to the
standard scenario.

\begin{figure}
  \centering
   \centering \begin{turn}{-90} \includegraphics[width=.333\textwidth]{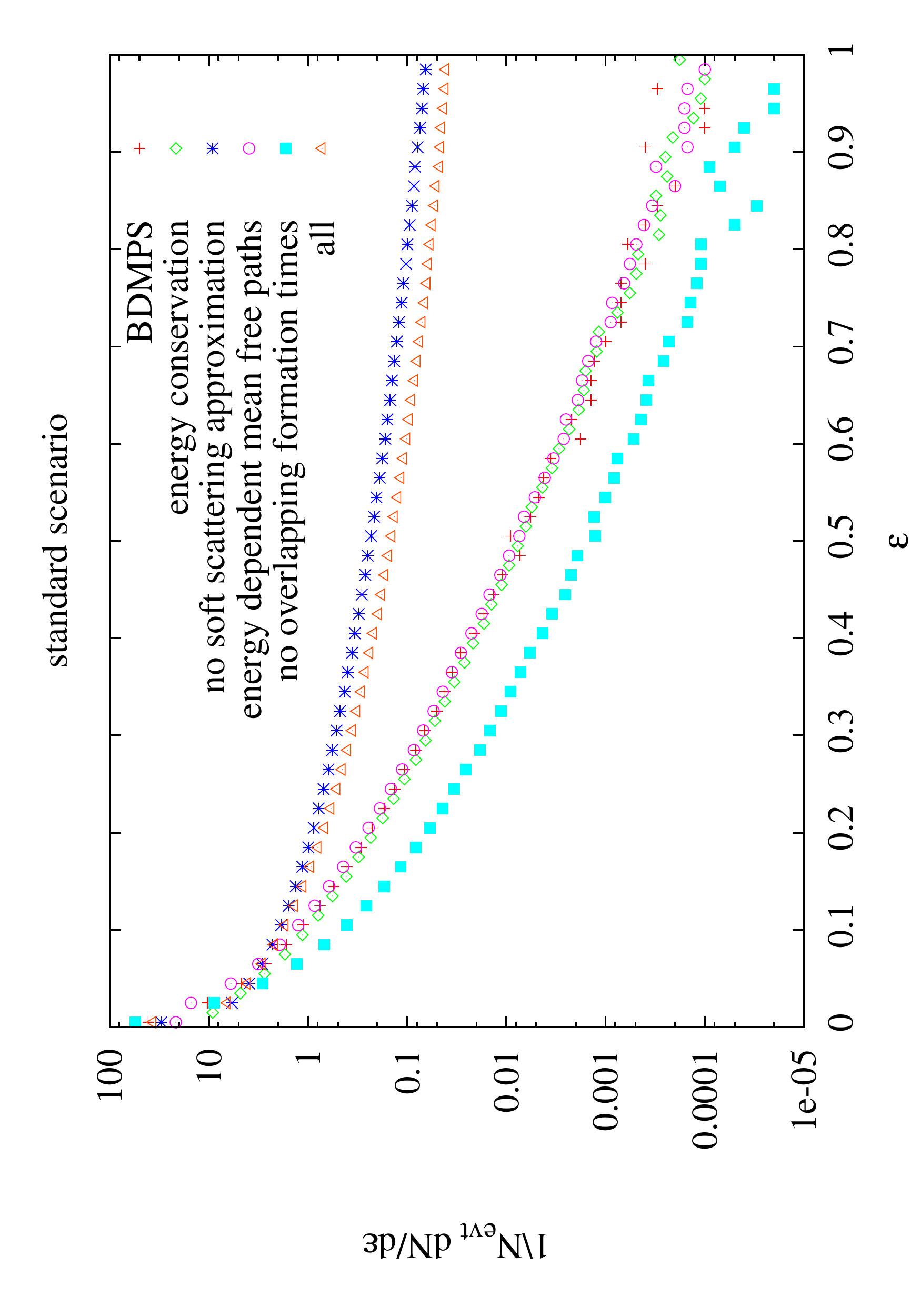} \end{turn}
   \centering \begin{turn}{-90} \includegraphics[width=.333\textwidth]{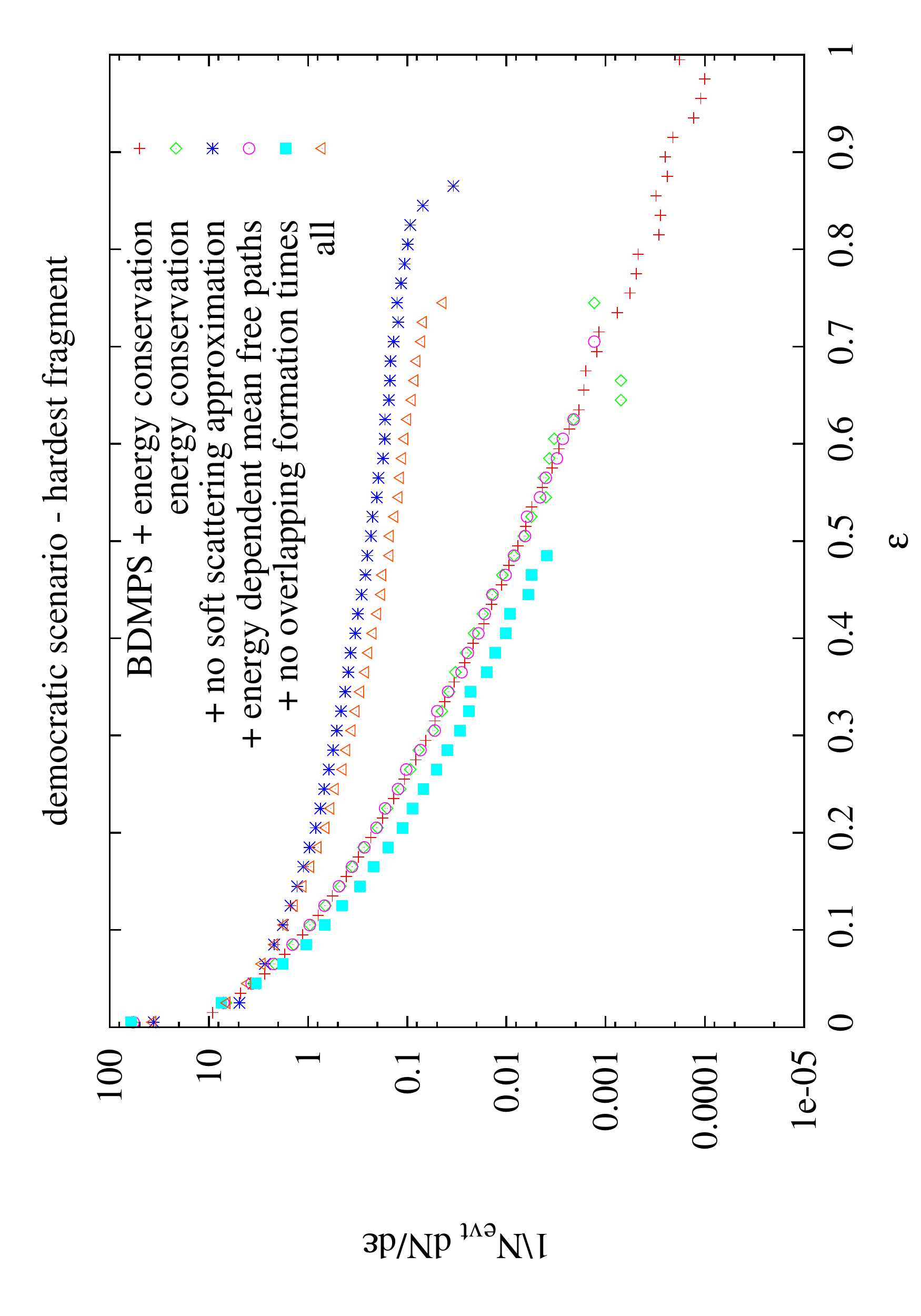} \end{turn}
   \caption{MC results for the distribution of the fractional energy loss $\epsilon = \Delta E / E_\text{proj}$ in the
standard (top) and democratic (bottom) scenarios for the parameters of Fig.~\ref{fig::omega}).
The different definitions of $\Delta E$ in both scenarios are
explained in the text. The probability of no parton energy loss has a discrete weight $p_0\, \delta(\epsilon)$
and is included in the first bin.  
 Note that in the
lower panel the 'BDMPS + energy conservation' calculation has the standard definition of the energy loss }
   \label{fig::de-dist}
\end{figure}

A more differential way of looking at the energy loss is by calculating the distribution of the energy loss at a fixed
path length $L$ as shown in Fig.~\ref{fig::de-dist}. The distributions all fall steeply with a rather significant
probability for no energy loss at all. In terms of differences between the different modifications the results 
reflect closely the findings discussed for the total energy loss. In the democratic scenario, virtually the only 
difference between the
two definitions of the energy loss is that the distributions with the energy loss derived from the most energetic
fragment at some point break off. This is most prominent in the case without overlapping formation times: Here the
distribution stops at $\epsilon=0.5$ because it is very unlikely to radiate more than one gluon per event. In both
scenarios,  removing the soft scattering assumption induces the largest modifications.

\begin{figure*}
  \centering
  \begin{minipage}[b]{.5\textwidth}
   \centering \begin{turn}{-90} \includegraphics[width=.7\textwidth]{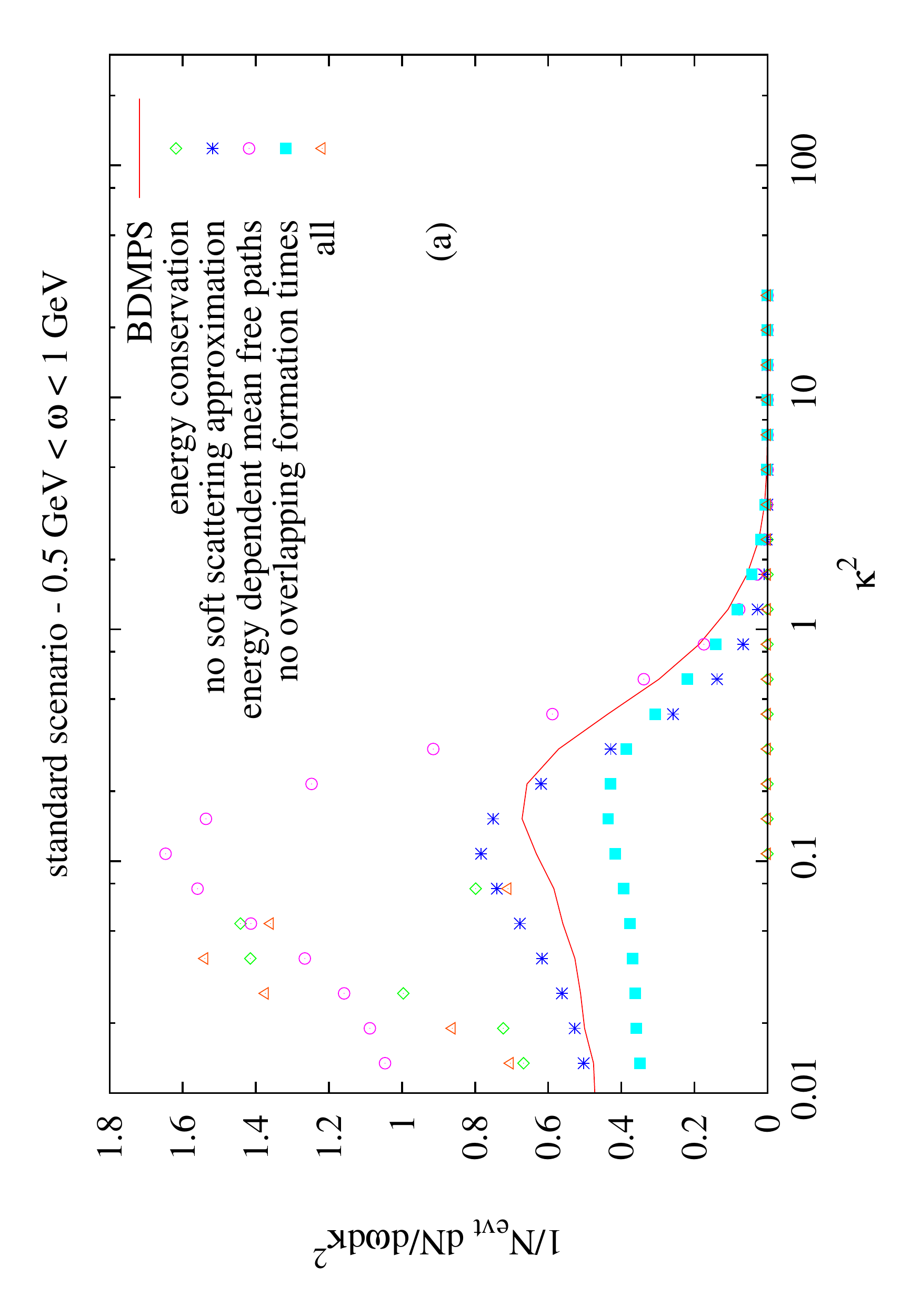} \end{turn}
  \end{minipage}%
  \begin{minipage}[b]{.5\textwidth}
   \centering \begin{turn}{-90} \includegraphics[width=.7\textwidth]{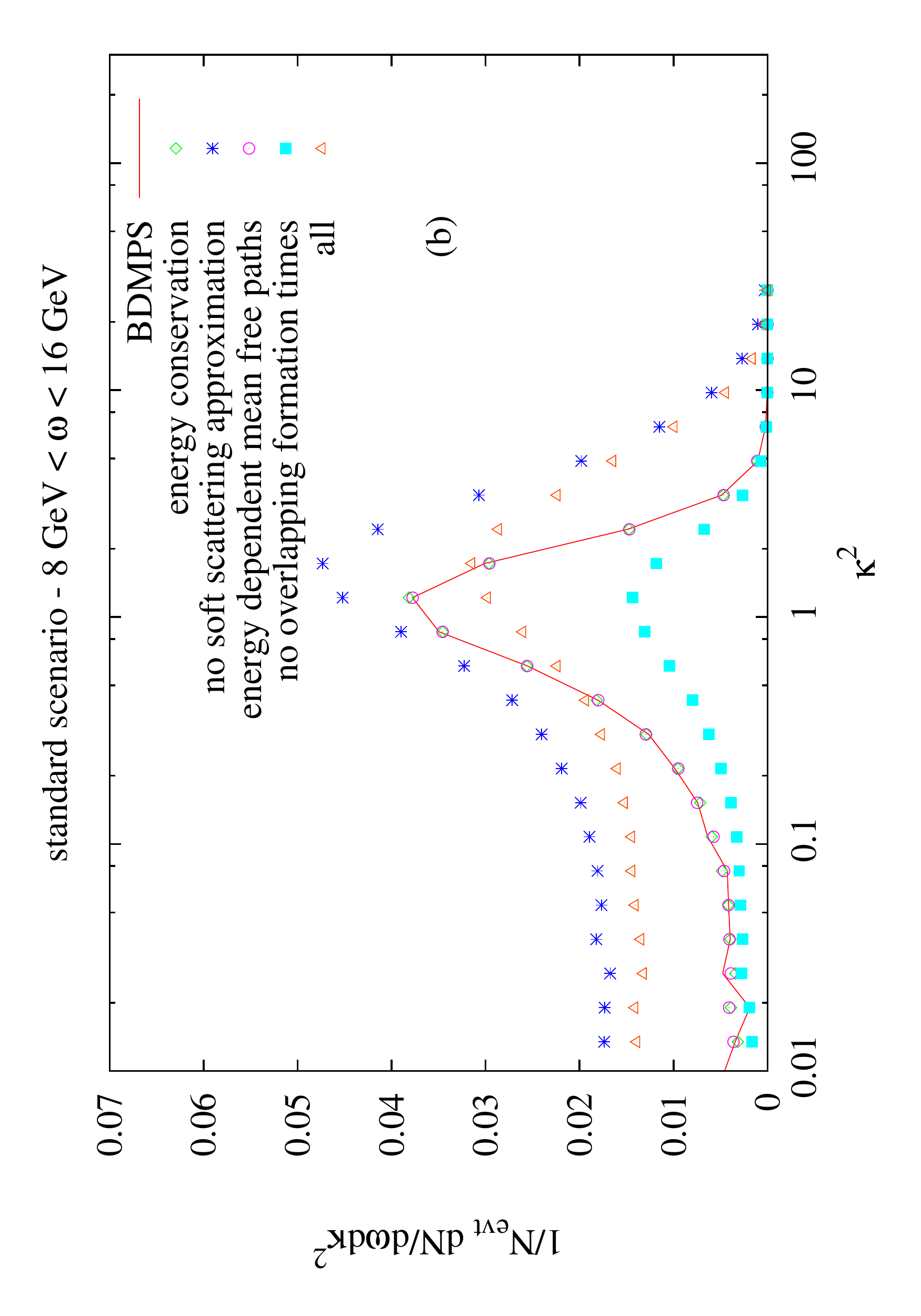} \end{turn}
  \end{minipage}\\
  \begin{minipage}[b]{.5\textwidth}
   \centering \begin{turn}{-90} \includegraphics[width=.7\textwidth]{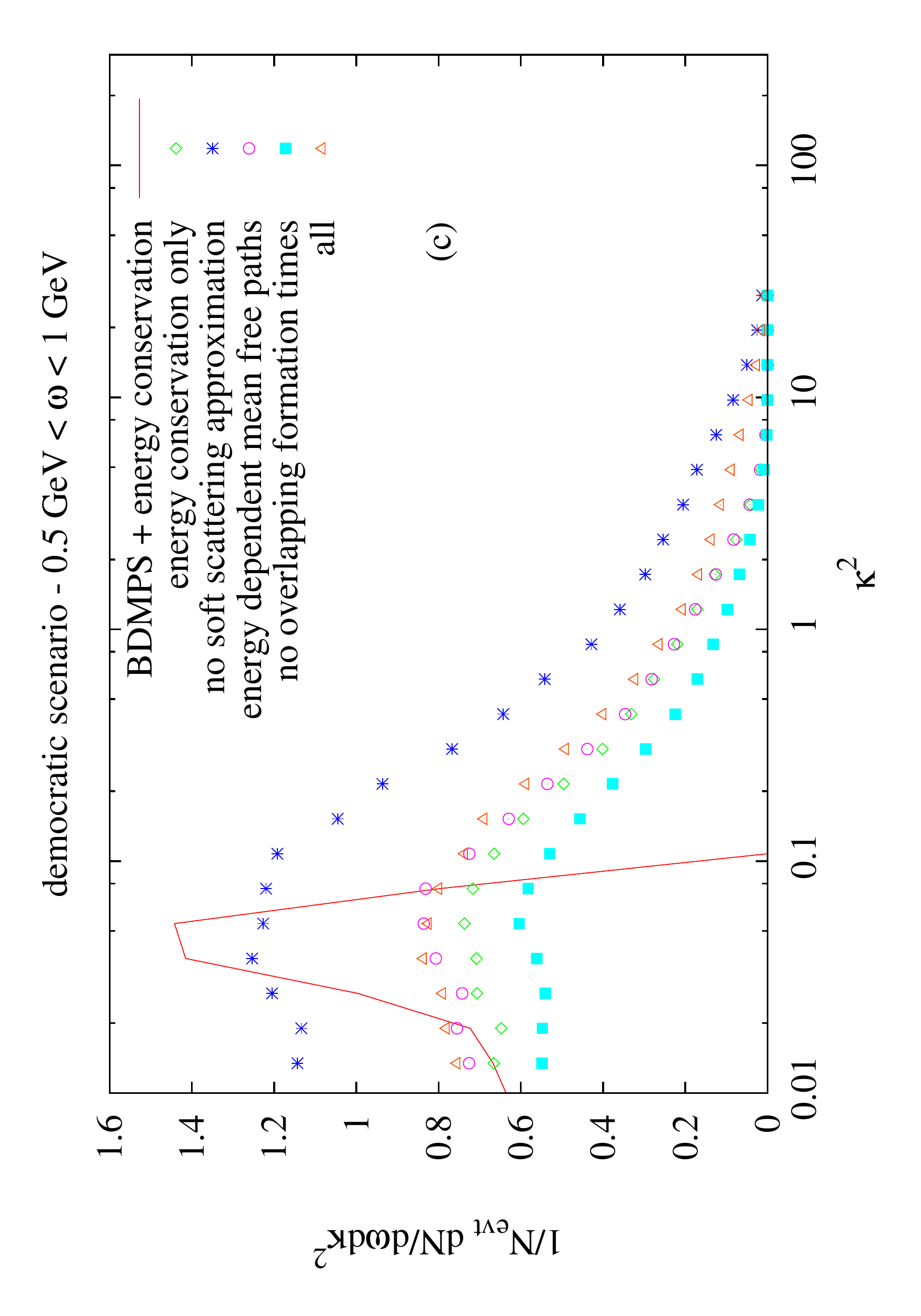} \end{turn}
  \end{minipage}%
  \begin{minipage}[b]{.5\textwidth}
   \centering \begin{turn}{-90} \includegraphics[width=.7\textwidth]{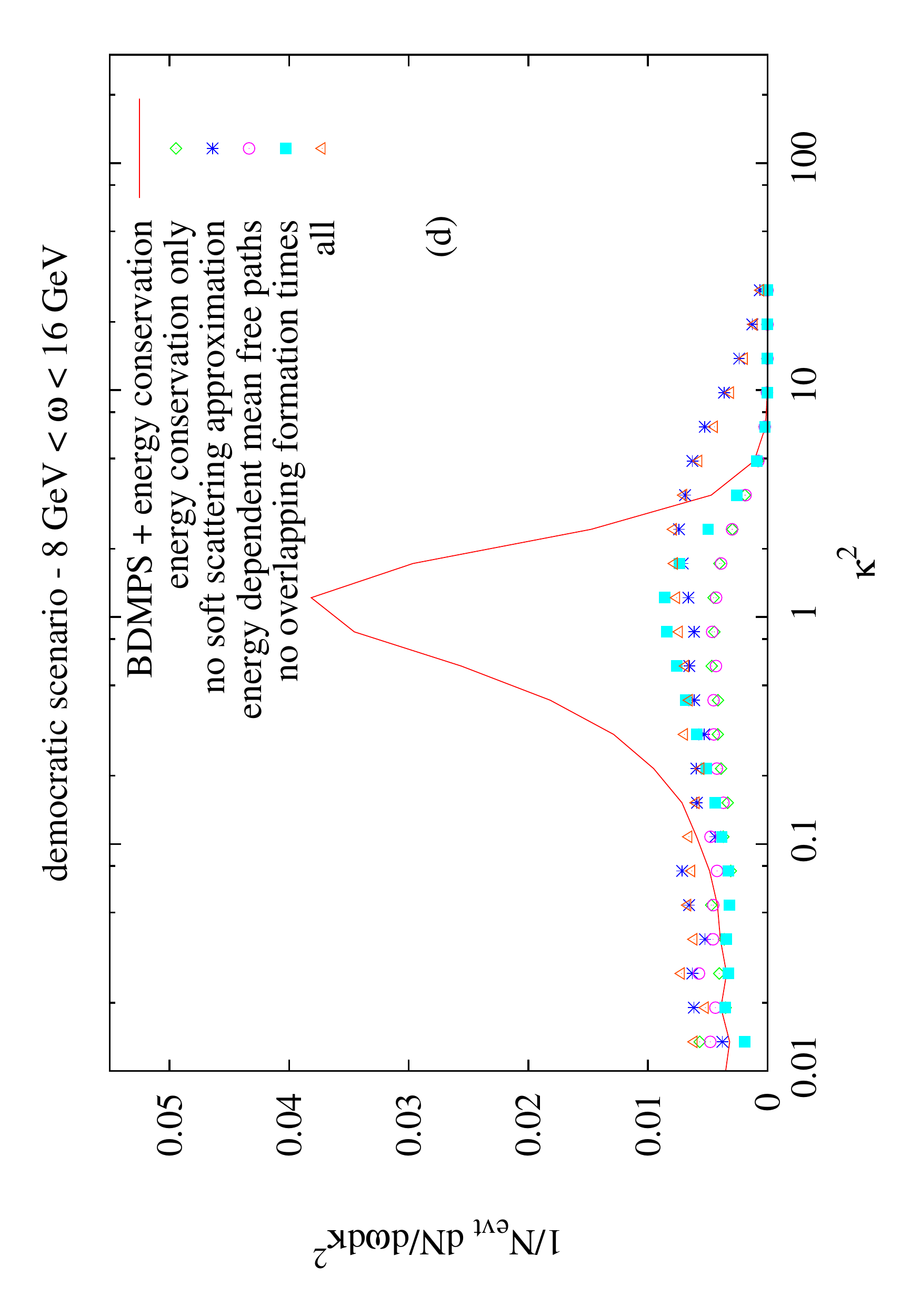} \end{turn}
  \end{minipage}\\
  \begin{minipage}[t]{\textwidth}
   \caption{MC results for the transerve momentum distribution ($\kappa^2 = \kt^2/(\qhat L)$) in one low and one high
gluon energy bin in the standard and democratic scenario (parameters as in Fig.~\ref{fig::omega}).}
   \label{fig::kappa}
  \end{minipage}
\end{figure*}

\section{Transverse momentum distribution}
\label{sec5}

Fig.~\ref{fig::kappa} shows the transverse momentum distribution of radiated gluons in two energy bins for the 
cases that subsequent branchings of radiated gluons are ('democratic scenario') or are not ('standard scenario')
allowed. The transverse momentum is defined relative to the incident projectile's direction. 

In the standard scenario, energy and momentum conservation cuts out the parts of the distribution with
$\qt > \omega$, as can be seen in the low energy bin (Fig.~\ref{fig::kappa}~(a)). At this gluon energy removing the
soft scattering constraint restricts the $\qt$ range, which shifts the transverse momentum distribution towards smaller
$\kt$ values. Forbidding formation times to overlap, on the other hand, leads to
an overall suppression without affecting the shape of the gluon distribution much. Finally, the sum of all effects is
dominated by energy-momentum conservation. This changes drastically at higher gluon energies
(Fig.~\ref{fig::kappa}~(b)). Here, conservation of energy and momentum and making the cross sections energy dependent
has no effect at all on the transverse momentum distribution. The assumption that formation times cannot overlap again
leads to a suppression but does not change the shape. In the case without soft scattering approximation, on the other
hand, the more energetic gluons can profit from the tail of the elastic scattering cross section and can acquire
significantly higher transverse momenta. This feature is also visible when all modifications are considered together. 

In the democratic scenario the most striking feature is that the distribution extends to much larger transverse momenta.
This is natural as in this scenario gluons radiated off gluons start out with a sizeable transverse momentum relative
to the projectile or jet axis. This effect becomes less pronounced at larger $\omega$, as energetic gluons are mainly
radiated off the projectile quark. In return another effect becomes visible, namely the smearing out of the peak which
has to do with the fact that gluons can radiate other gluons and thus move to lower energies. Again, the most 
significant modifications compared to the BDMPS-Z limit result from imposing energy and momentum
conservation and from removing the soft scattering approximation for interactions
between jet and medium.

\section{Summary and Conclusions}
\label{sec6}

The present study documents for exemplary cases how elementary 
physical requirements (such as energy-momen\-tum conservation or the democratic treatment of all scattering processes)
and different physical assumptions about the jet-medium interaction can affect the simulation of jet quenching phenomena.
We emphasize that the significant differences between the model cases explored here should not be viewed 
as systematic uncertainties of jet quenching simulations. Rather, energy-momentum conservation and a dynamical
treatment of all scattering and branching processes on an equal footing are clearly sensible physical requirements. 
To the extent to which imposing these requirements changes the result of a simulation, a simulation without them 
should not be regarded as reliable. This prompts us to conclude that 
these requirements are prerequisites
if one wants to employ Monte Carlo simulations of parton energy loss for separating between different pictures
of jet-medium interactions, such as the two different pictures described in section~\ref{sec::steps}.  Our study also further supports the
argument~\cite{Armesto:2011ht} that the BDMPS-Z formalism (that does not impose exact energy-momentum conservation and 
that is restricted to describing the medium-induced radiation of a nominally leading parton) cannot be regarded as a phenomenologically
viable approximation for the full dynamics relevant for jet-medium interactions, although it is an interesting, analytically accessible 
limiting case of the full dynamics of parton energy loss. 

The present study also explains how one can understand qualitatively (and to some extent even quantitatively) the 
deviations from the BDMPS-Z parton energy loss in the 
high-energy limit (\ref{eq::bdmps}) induced by imposing the physical requirements listed in section~\ref{sec::steps}. In particular, the 
role of medium-induced destructive interference depends sensitively on the nature of the jet-medium interaction. 
Harder momentum transfers from the medium can destroy coherence on shorter length-scales and result in
quasi-incoherent inelastic interactions with enhanced energy loss  and a linear dependence of the average parton
energy loss on the medium length. This strong sensitivity is of significant interest as it may help to constrain the
nature of the jet-medium interactions. Our discussion in sections~\ref{sec3}-\ref{sec5} has also identified simple
physical origins for the effects resulting from supplementing the results of the BDMPS-Z formalism with
exact energy-momentum conservation, democratic
treatment of all partons and no overlap between gluon formation times. 

While our study illustrates the insufficiency of a parton energy loss calculation based on the 
eikonal limit (\ref{eq::bdmps}) alone, we note that the extension of eikonal kinematics to the full kinematic range is not
unique. Thus our study cannot provide a {\it unique} solution to the question of how the full kinematical range
relevant for jet-medium interactions should be modeled. 
To achieve a formulation valid in the entire phase space, one may want to start therefore from a formulation that 
is consistent with the known results in the eikonal limit, but that does involve neither eikonal approximations 
nor concepts that are only meaningful in eikonal kinematics. Such an approach is taken for instance in a new
version of the medium-modified final state parton shower \textsc{Jewel} (for first results, see 
Ref.~\cite{Zapp:2011ek}) that aside of other phenomenologically wanted features also accounts for the 
physical requirements listed in section~\ref{sec::steps}. More generally, we hope that the present study contributes
to clarifying those physical features that a phenomenologically viable MC simulation of parton energy loss must
satisfy.


\end{document}